# High-order harmonics with frequency-varying polarization within each harmonic

Avner Fleischer[1,2], Ofer Kfir[1], Pavel Sidorenko[1], and Oren Cohen[1]

[1]Solid State Institute and Physics department, Technion, Haifa, 32000, Israel
[2]Department of Physics and Optical engineering, Ort Braude college, Karmiel, 21982, Israel.

**Abstract**

We predict high-order harmonics in which the polarization within the spectral bandwidth of each harmonic varies continuously and significantly. For example, the interaction of counter-rotating circularly-polarized bichromatic drivers having close central frequencies with isotropic gas leads to the emission of polarization-fan harmonics which are nearly circularly-polarized in one tail of the harmonic peak, linear in the center of the peak and nearly circular with the opposite helicity in the opposite tail. Polarization fan harmonics are obtained as a result of multiple (at least two) head-on recollisions of electrons with their parent ions occurring from different angles. The process can be phase-matched using standard methods (e.g. pressure tuning phase matching) and maintains the single-atom polarization property through propagation. These polarization-fan harmonics may be used for exploring non-repetitive ultrafast chiral phenomena, e.g. dynamics of magnetic domains, in a single shot.

High harmonic generation (HHG)-based sources have been used for the past 25 years for a variety of applications in ultrafast spectroscopy and high resolution imaging of atoms, molecules and nano-structures [1-6]. As such, they are the subject of intensive study. However, while many schemes have been developed for controlling the spatial, temporal and spectral properties of HHG-based light sources [7-11], the polarization degree of freedom of HHG radiation was until recently largely uncontrollable. Moreover, in all previous theoretical and experimental works on HHG, the polarization within each harmonic was constant. The possibility to produce high-order harmonics in which the polarization varies spectrally within a single harmonic profile (peak) has never been studied or even considered.

The difficulty in controlling the polarization of the emitted harmonics is associated with the fact that HHG is a recollision phenomenon [12-13] whose efficiency is highest for "head-on" recollisions with the parent ion, which for atomic media and monochromatic driver is induced by linearly-polarized driver only. In this case symmetry dictates that the polarization of the harmonics obtained would be linear as well. Indeed, this is the geometry used in the vast majority of HHG experiments. Introducing slight ellipticity into the driver results in the emission of HHG with relatively small ellipticity (<0.4) [14-16] with efficiency that decreases rapidly with increasing ellipticity of the driver [17-18]. When driven by circularly polarized laser field however, the electron completely misses the parent ion; hence high harmonics are not emitted. While the experimental progress in controlling the polarization of high harmonics has been very limited for many years [19-22] ,many ideas were proposed for overcoming this challenge [23-32]. Recently, HHG with fully tunable polarization, from circular through linear to circular polarization in the opposite helicity, was demonstrated [33]. Phase matching of circularly (but not elliptically) -polarized HHG and their downstream utilizations were also demonstrated [34]. These works made polarization a new controllable property in HHG. Still, the capability for varying and controlling the polarization within the harmonic peak is missing.

Here, we propose a completely new concept in HHG: ellipticity-fan HHG, where the polarization within each harmonic peak is not fixed but rather frequency-dependent. For example, it can vary continuously all the way from nearly-circular with left helicity, through linear to nearly-circular with right helicity. These ellipticity-fan harmonics are generated through wave-mixing of two co-propagating driving laser pulses with slightly different central-frequency and circular polarization with opposite helicities in isotropic gas. Importantly, this process can be phase matched efficiently by using standardly applied phase-matching techniques of linearly polarized HHG that are driven by linearly polarized quasi-monochromatic driver. The proposed ellipticity-fan high harmonics may be used for circular dichroism experiments without periodically alternating the helicity of the probe beam, as is currently done in this type of studies. Consequently, it may be used for single-shot experiments, e.g. for exploring non-repetitive phenomena (where pump-probe is inadequate) such as dynamics of magnetic domains.

Our method can be understood by examining the selection rules of HHG driven by counter-rotating bi-chromatic drivers of arbitrary frequencies, $\omega_1$ and $\omega_2$. In this case, the allowed emitted harmonic frequencies are given by [33,35]:

$$\Omega_{(n_1,n_2)} = n_1 \cdot \omega_1 + n_2 \cdot \omega_2 \quad , \quad n_1 + n_2 \in 2k-1 \qquad (1)$$

where $n_1$ and $n_2$ are integer numbers that can be associated with the number of 'driver photons' annihilated in the process from the drivers at angular frequencies $\omega_1$ and $\omega_2$, respectively. Parity conservation requires that $n_1+n_2$ is odd (k is an integer). Each high harmonic frequency $\Omega_{(n_1,n_2)}$ can in principle be identified by the pair $(n_1,n_2)$. When the polarizations of the first and second colors are respectively left- and right-circular then conservation of spin angular momentum dictates that only two possible harmonic channels exist: ($n_1$=n+1,$n_2$=n) left circularly-polarized harmonics, and (n,n+1) right circularly-polarized harmonics (i.e. consecutive harmonics exhibit opposite helicity). The underlining principle behind our scheme is that by taking the central frequencies of the bi-chromatic drivers close enough to each other, every pair of consecutive harmonics, (n+1,n) and (n,n+1) merge into a single harmonic peak. Assuming $\omega_1<\omega_2$, the low-frequency tail of each of these harmonics is dominated by the (n+1,n) channel and therefore exhibit nearly left-circular polarization while the high-frequency tail is dominated by the (n,n+1) channel and exhibit right-circular polarization. The polarization is linear in the harmonic's center where the channels overlap evenly.

We have explored the generation of HHG with varying polarization using numerical simulations. We assume bi-chromatic counter-rotating drivers with central frequencies $\omega_1 = \omega \cdot (1-\delta)$ and $\omega_2 = \omega \cdot (1+\delta)$ where ω corresponds to a wavelength of 800 nm and $\delta = (\omega_2 - \omega_1)/(\omega_2 + \omega_1) > 0$ is the normalized frequency difference. The polarization of each driver is controlled independently by a quarter wavelength wave-plate (WP). The bi-chromatic pulse $E_{BC}(t)$ is a superposition of the two driver pulses, $E_{BC}(t)=E_1(t)+E_2(t)$ that are given by:

$$E_i(t) = \sqrt{I_{0,i}} g(t) \left\{ (-1)^i \sin\alpha_i \cos\alpha_i \left[\cos(\omega_i t) + \sin(\omega_i t)\right] \mathbf{e_x} + \left[\cos^2\alpha_i \cos(\omega_i t) - \sin^2\alpha_i \sin(\omega_i t)\right] \mathbf{e_y} \right\} \quad (2)$$

where i=1,2 correspond to the first and second field, respectively. In Eq. (2), $I_{0,i}$ is the pulse peak intensity, $g(t) = \exp\left[-2\ln 2 (t/\tau)^2\right]$ is a Gaussian envelop with width (full width at half maximum) of $\tau = 13.35 fs$ and $\alpha_i$ is the angle of the WP's fast axis with respect to the polarization axis of the incoming pulse ($\alpha_1$=45$^0$ and $\alpha_2$=45$^0$ result with left and right circular polarization, respectively, i.e. circular polarization with helicity of h=+1 and h=-1, respectively). Notably, such bi-chromatic counter-rotating drivers can be produced experimentally using several approaches, including i) launching a quasi-monochromatic linearly-polarized ultrashort laser pulse into a Mach–Zehnder-like interferometer where dichroic mirrors with a step transmission function at 800nm are used as beam splitters and combiners, ii) using polarization pulse shaping techniuqes and iii) optical parametric amplifiers. We integrate the three-dimensional time-dependent Schrödinger equation (3D TDSE) for a single electron in a model potential for Argon atom $V(\mathbf{r}) = -(1 + 0.2719 e^{-0.25 \mathbf{r} \cdot \mathbf{r}}) / \sqrt{0.09192 + \mathbf{r} \cdot \mathbf{r}}$, where **r** is the coordinate vector, interacting with the bi-chromatic pulse, $E_{BC}(t)$, (for more details see [31]). The dipole acceleration expectation value, which is proportional to the emitted classical field, is calculated using the time-dependent wavefunction. The two components of the dipole acceleration Fourier transform are used for calculating the HHG spectra (HGS) and the polarization of the harmonics.

We first present in Fig. 1 polarization-fan harmonics in which the ellipticity varies continually from nearly left circular polarization, through linear polarization to nearly right circular polarization. In this calculation, the two driver pulses are circularly-polarized ($\alpha_1=\alpha_2$=45$^0$). The peak intensities are $I_{0,1} = I_{0,2} = 1.12 \cdot 10^{14} W/cm^2$. Figure 1a shows the HGS as a function of the normalized frequency difference $\delta$. When $\delta$ is large, consecutive spectrally-separated harmonic peaks are generated. According to the photonic picture, a pair of integers could be assigned for each harmonic peak which expresses the number of photons absorbed from each driver pulse. For instance, channel (7,8) appears at frequency $\Omega_{(7,8)} = 7\omega(1-\delta) + 8\omega(1+\delta) = \omega \cdot (15+\delta)$ and is right-circularly-polarized and channel (8,7) appears at $\Omega_{(8,7)} = \omega \cdot (15-\delta)$ and is left-circularly-polarized. As $\delta$ decreases, the harmonic channels (n+1,n) and (n,n+1) approach each other and merge into composite harmonic channels. This is evident from Fig. 1b,d which show HGS lineouts of Fig.1a around the 9$^{th}$ and 15$^{th}$ harmonics, for progressively decreasing values of $\delta$. In the limit δ=0 the resultant driver field is linearly polarized, and the two channels coalesce to a single harmonic peak at the odd frequency (2n+1)ω. Figures 1c,e show the elipticity-helicity product for the 9$^{th}$ (n=4) and 15$^{th}$ (n=7) harmonics, for decreasing values of $\delta$. Rather than obtaining strictly linearly- or elliptically-polarized harmonics [33,34] (where the polarization was fixed within each harmonic), here each harmonic peak spans a fan of polarization states, ranging from right highly-elliptical (at the low-energy wing of the peak), through linear (at the peak's maxima) to right highly-elliptical (at the other wing). Being the result of the mergence of two circularly-polarized harmonics with alternating helicities, the polarization properties of its "parent" harmonics are maintained only at the peak wings, where at the center of the peak the two circular polarization states average up to linear polarization. As $\delta$ decreases the transition from one helicity state at one wing of the harmonic peak to the opposite helicity state at the other wing becomes shallower. In order to gain time-domain perspective for the polarization-fan, Figs.1f,g show bi-chromatic driver pulse for $\delta = 0.1$ and $\delta = 0.025$, respectively. The driving field is nearly linearly-polarized with its polarization axis slowly rotating. This field induces exactly two recollision events per optical cycle $T = 2\pi/\omega$, each having a "head on" character in attacking the parent ion, forming an angle of $\pi(1-\delta)$ between their attack directions. We now show that this discrete rotation of the attack direction gives rise to the polarization fan. To this end, we calculated the spectral field, $\tilde{E}_{APT}(\Omega)$, of an attosecond pulse train (APT) of 8 linearly-polarized delta-function bursts where the polarization direction rotates by $\pi(1-\delta)$ from one burst to the next

$$\tilde{E}_{APT}(\Omega) = -\sum_{n=0}^{7} \exp\left(-i\Omega n \frac{T}{2}\right)\left\{\cos\left[\pi(1-\delta)n\right]\mathbf{e_x} + \sin\left[\pi(1-\delta)n\right]\mathbf{e_y}\right\} \quad (3)$$

Where $\mathbf{e_x}, \mathbf{e_y}$ are unit vectors in the x- and y-directions, respectively and the first burst (at time zero) was taken to attack the ion in the negative x direction. The ellipticity is calculated from the x- and y- components of the spectral field. As shown in Fig. 1e (dashed curves), the calculated polarization of the harmonics is indeed fanned, with fairly good matching to the TDSE calculation, proving that the polarization fan results from the nearly π rotation in the attacking angle of consecutive recollisions.

Having established that the polarization fan relies on synthesis of several recollisions, we now explore how its properties depend on the number of recollisons and what the minimum number of required re-collisions is. Thus, we simulated a polarization gating setup [36] where the delay (overlap) between two ultrashort ($\tau = 5.34\,fs$) circularly-polarized counter-rotating pulses in a (0.9ω, 1.1ω) scheme was controlled. Figure 2 shows HHG fields (harmonics 13-24) for several different delays. When the two pulses fully overlap, an APT consisting of 4 bursts is obtained, each emitted along a slightly different direction (Fig. 2a). The harmonic intensity profile and ellipticity-helicity product are shown in Figs. 2e and 2f, respectively. Decreasing the overlap between the two pulses reduces the number of recollisions (Figs. 2b-d), the harmonics intensity (Fig. 2e) and the ellipticity at the harmonic wings (Fig. 2f). Still, even two re-collisions can give rise to polarization fan (Figs. 2d,f). That is, polarization fan high harmonics (of 800 nm driver) can correspond to HHG fields with nearly femtosecond pulse duration, a possibly important feature for future applications.

We presented generation of high harmonics with ellipticity fan. Next, we present high harmonics with more complex polarization structure by using bi-chromatic drivers in which one or both drivers are elliptically polarized. Figure 3 presents a case in which $\delta = 0.1$ and the first color is elliptical with ellipticity $\varepsilon_1 = tg(\alpha_1)$ while $\alpha_2=45^0$. This symmetry breaking leads to generation of new harmonic channels (11,4), (10,5), (9,6),.., (6,9), (5,10),... [33,37]. The harmonic channels are marked by the yellow indices ($n_1,n_2$) in Fig.4a which shows the HGS vs. the reading $\alpha_1$ of the first waveplate. The 9th-harmonic line-shape and ellipticity-helicity product, $\varepsilon \cdot h$, are presented in Figs. 3b and 3c, respectively for several values of $\alpha_1$. Figs. 3d and 3e display the same information for the 15th-harmonic. Clearly, the polarization structure of these exemplary harmonics depends on $\alpha_1$. For example, the polarization of the central part of the peak can be tailored from right-elliptical, through linear, to left-elliptical by changing the reading of $\alpha_1$ by as little as 6 degrees in either direction away from $\alpha_1=45^0$. The strong dependence of the polarization structure on $\alpha_1$ results from the overlap of several channels (and not just two, as in the symmetric case of $\alpha_1 = 45^0$) which coalesce into a single harmonic peak with complicated polarization fan. Finally, we present high harmonics with modulated ellipticity in Fig. 4. In this case, we assume bi-chromatic drivers that were generated by splitting a single Gaussian laser pulse centered around 800nm into two parts by a dichroic beam splitter with a sharp edge transmission curve at 800nm. The initial linear pulse supports 10 optical cycles ($\tau = 26.7\,fs$) and the two pulses are made elliptically-polarized and counter-rotating ($\alpha_1=16.5^0$, $\alpha_2=73.5^0$). The frequency dependence of the polarization within the 15$^{th}$ and 21$^{st}$ harmonic peaks are shown in Figs. 4c and 4e, respectively. The modulated ellipticity results from the fact that every harmonic peak consists of very many new channels with different ellipticities. This can also be explained from a time-domain perspective. Figure 4f shows that the bichromatic driver field is elliptically-polarized most of the time, i.e. it does not induce recollisions, except for two periods in which it is very close to linear: at t~93.6T and t~105.9T. The two groups of recollsions [see Figs. 4g] which occur at those periods interfere, and are responsible for the spectral modulations appearing in both the HGS and the ellipticity-helicity product, [see Figs. 4(b-e)]. Indeed, the period of these modulations is $\Delta\Omega = 2\pi/12.3T \cong 0.081\omega$, as predicted theoretically.

Before concluding, it is noteworthy to mention two essential properties of our proposed process. First, we have verified that the polarization-fan is largely insensitive to variations in the driver's intensities and intensity ratio and hence survives volume averaging in the focus (see Supplementary Material). Second, it can be phase-matched using standard phase-matching methods in HHG (e.g., pressure-tuning phase matching) [38], due to the very small dispersion between the two drivers which have very close frequencies. For comparison, phase matching of elliptically polarized high harmonics has never been studied before, while that of circularly polarized HHG was demonstrated only in Ref. [34]. Thus, we anticipate that experiments would yield bright and robust polarization-fan harmonics.

In conclusions, we propose and demonstrated numerically polarization-fan high harmonics: harmonics in which the polarization varies significantly within the spectral bandwidth of each harmonic-order. Generation of harmonics with polarization that varies from nearly circular polarization with left helicity through linear to nearly right circular polarization is presented. Notably, the process can be phase matched, hence it fills a currently existing gap in producing bright (phase-matched) elliptically-polarized high harmonics. This source may be useful for applications. Polarization should soon become an accessible useful knob in HHG experiments [33,34], including in HHG spectroscopy and in downstream experiments. Importantly, experiments with chiral light often require fast alternation of the light source between right and left helicities during the measurement (e.g. x-ray magnetic circular dichroism signals correspond to the normalized difference between the transmission or reflectance of the right and left circularly polarized light). As the polarization-fan harmonics contains both helicities at near frequencies, it may remove this requirement, simplifying these experiments. Also, it can potentially allow measurements of chiral process in a single shot, opening the door for measuring ultrafast non-repetitive chiral processes (i.e. where pump-probe techniques cannot be used) such as the demagnetization dynamics of a specific magnetic domain after an impulse excitation.

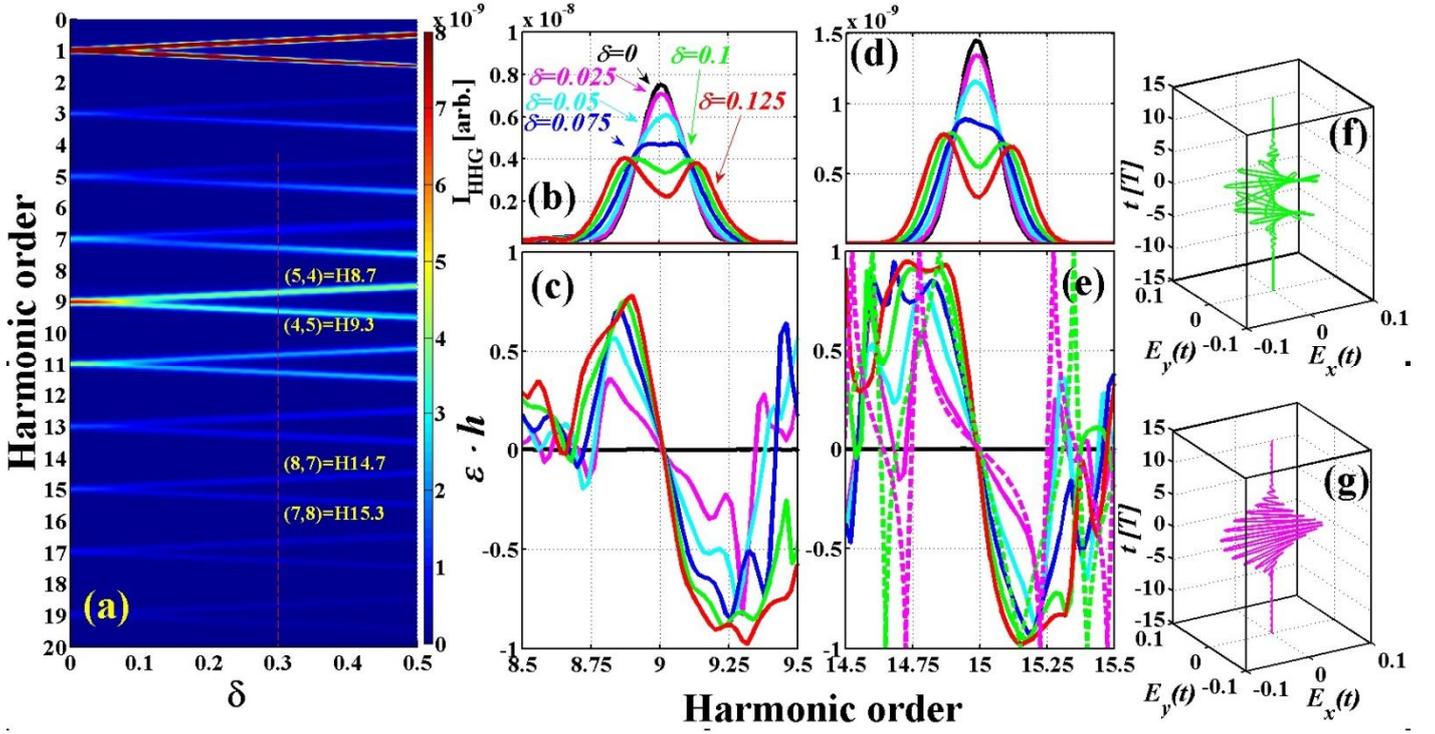

Figure 1: Polarization fan high harmonics. (a) HGS when the two driver pulses have central frequencies $\omega(1-\delta)$ and $\omega(1+\delta)$, as function of $\delta$. The two pulses are circularly-polarized and counter-rotating ($\alpha_1=\alpha_2=45^0$). The vertical dotted red line marks the value of $\delta=0.3$ for which the harmonic channels (marked by yellow) (8,7) and (7,8) appear in the HGS at harmonics 14.7 and 15.3, respectively. (b) a two-dimensional cut of (a) for the 9$^{th}$ harmonic, for several values of $\delta$: $\delta$=0.125,0.1,0.075,0.05,0.025,0 (red, green, blue, cyan, magenta, black curves, respectively). (c) ellipticity-helicity product $\varepsilon \cdot h$ for the 9$^{th}$ harmonic, for several values of $\delta$. (d)- same as (b) but for the 15$^{th}$ harmonic. (e)- same as (c) but for the 15$^{th}$ harmonic. The dashed lines are the results of the theoretical model (Eq. 3). Plots (f) and (g) display the total driver field for $\delta=0.1$ and $\delta=0.025$, respectively.

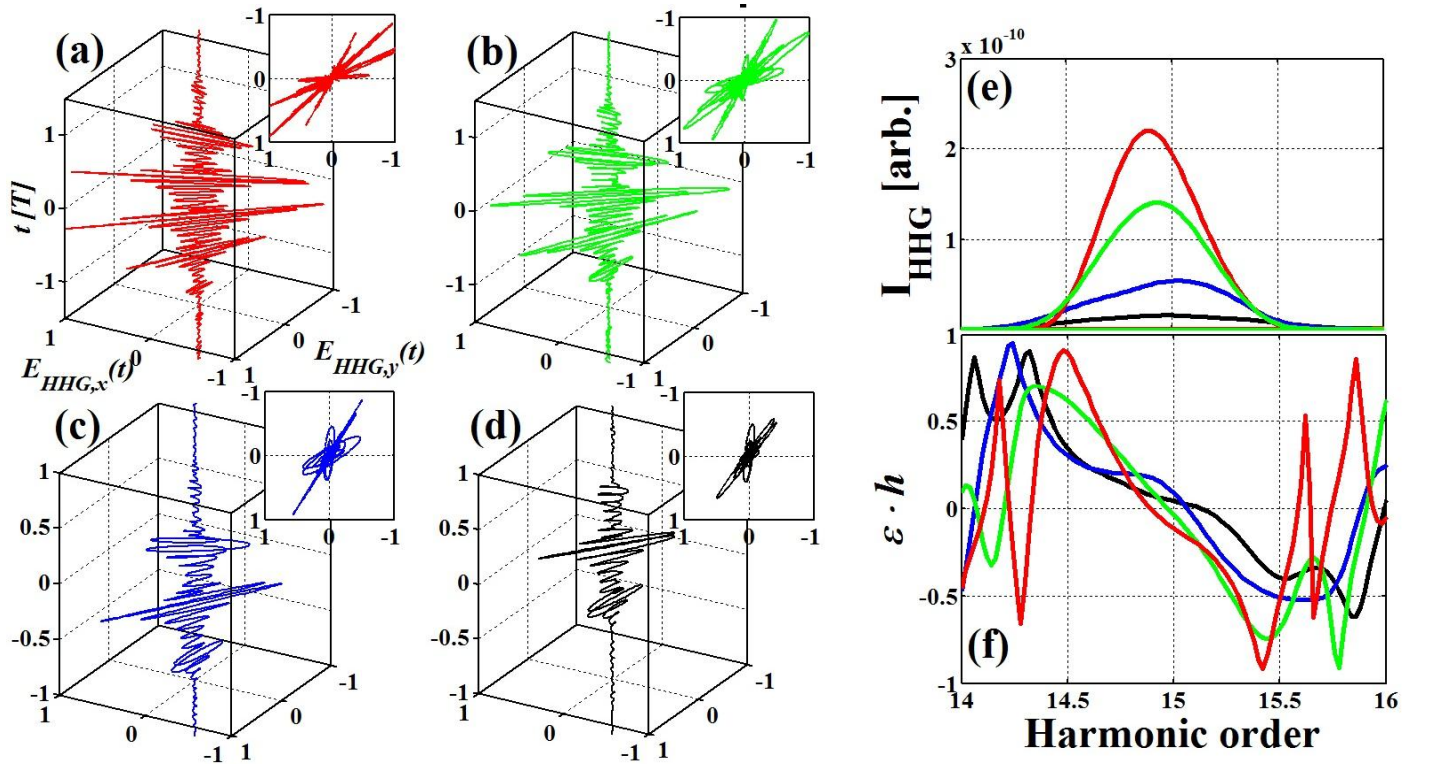

Figure 2: Polarization fan APT using polarization gating: (a-d) Vectorial time-dependent HHG field (harmonics 13-24) for a (0.9ω, 1.1ω) scheme ($\delta = 0.1$) where the two pulses are circularly-polarized and counter-rotating ($\alpha_1=\alpha_2=45^0$) and have duration of $\tau = 5.34\, fs$ (roughly 2 optical cycles). When the two pulses fully overlap an APT consisting of 4 bursts is obtained (a). Decreasing the overlap between the two pulses reduces the number of recollisions (Figs. 2b-d), the harmonics intensity (Fig. 2e) and the ellipticity at the harmonic wings (Fig. 2f). An APT comprising of just 2 recollisions still maintained the polarization fan property.

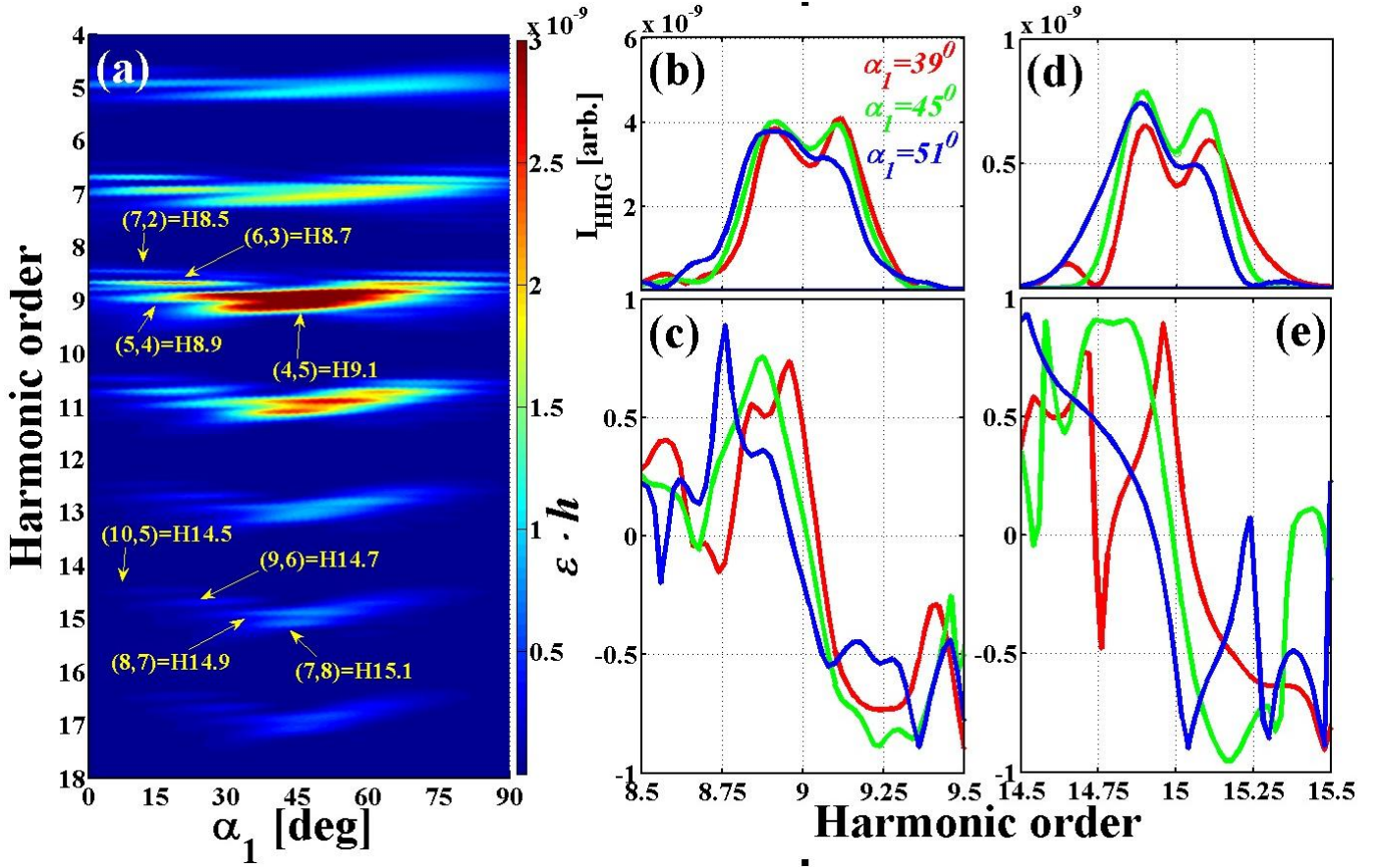

Figure 3: (a) HGS for a (0.9ω, 1.1ω) scheme ($\delta = 0.1$) where the second field is kept right circularly-polarized ($\alpha_2=45^0$), the polarization of the first color varies by scanning the reading $\alpha_1$ of its quarter waveplate and the two fields are counter rotating. Harmonic channels are marked by the yellow pairs ($n_1, n_2$). (b) a two-dimensional cut of (a) for the 9th harmonic, for several values of $\alpha_1$: $\alpha_1 = 39^0, 45^0, 51^0$ (red, gree, blue curves, respectively). (c) ellipticity-helicity product $\varepsilon \cdot h$ for the 9th harmonic, for several values of $\alpha_1$. (d)- same as (b) but for the 15th harmonic. (e)- same as (c) but for the 15th harmonic.

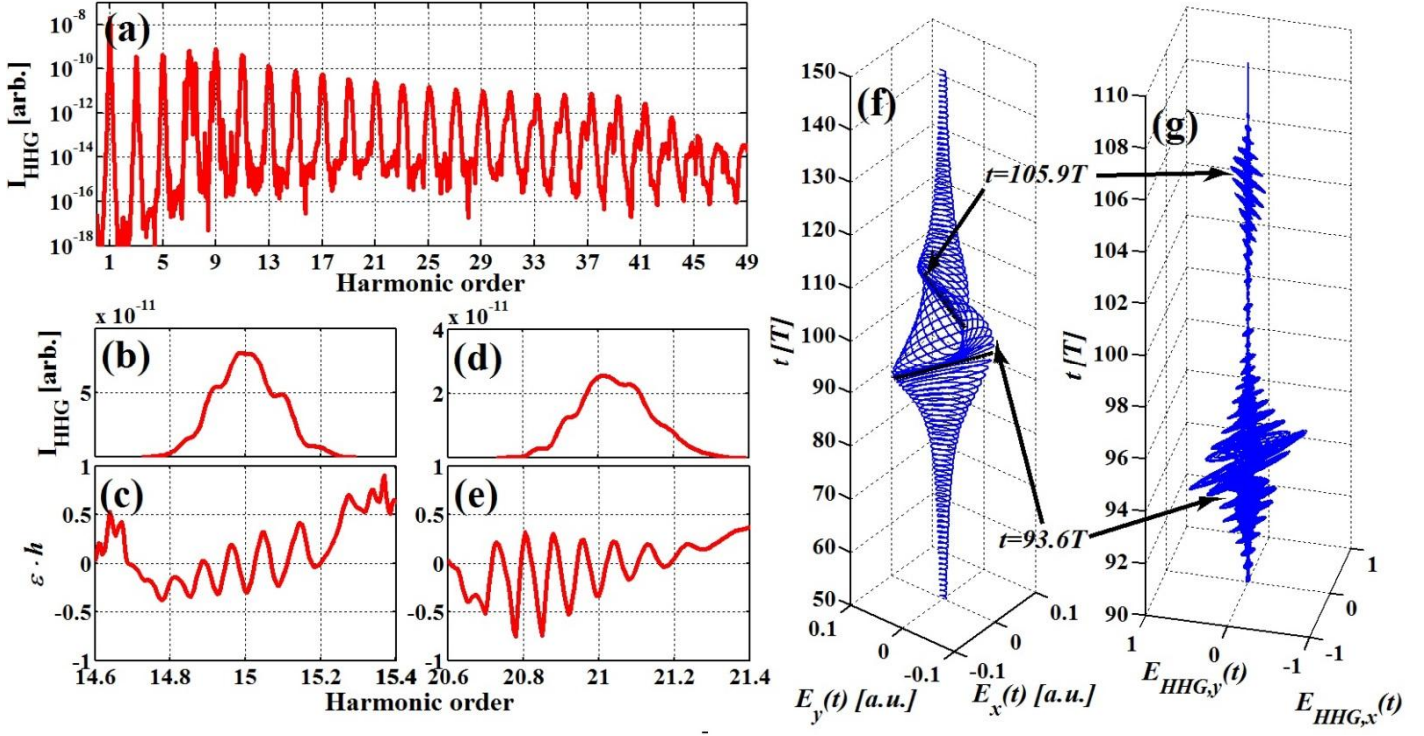

Figure 4: Harmonics with modulated ellipticity: the driver fields are obtained by using sharp edge dichroic mirrors. $\alpha_1=16.5^0$ and $\alpha_2=73.5^0$ yield right-elliptically-polarized driver field with wavelengths >800nm and left-elliptically-polarized driver field with wavelengths <800nm. Here many harmonic channels coalesce into every harmonic peak, giving a much more complex polarization profiles. (a,b,d) HGS, (c,e)- ellipticity-helicity product. Vectorial time-dependent driver field (f) and HHG (harmonics 13-45) field (g). The modulated polarization result from interference between two main attosecond bursts at t~93.6T and t~105.9T. In these periods the combined driver field is linearly polarized (see black arrows) while it is elliptically polarized in other times, hence suppressing the HHG emission.

*Corresponding author.

avnerf@tx.technion.ac.il

†Corresponding author.

oren@technion.ac.il

**Supplementary Material**

**Robustness of polarization fan high harmonics generation to volume averaging.**

HHG-based light sources usually rely on the focusing of the driver fields in one of 3 focusing geometries: a supersonic gas jet, a hollow-core fiber or a semi-infinite gas cell. The macroscopic HHG field is a coherent sum of emissions from atoms that experience different driver intensities and intensity ratios in the focusing volume, both along the propagation direction and in the transverse plane. Since we have established that the mechanism of polarization fan harmonics rely on the synthesis of several recollisions, their absolute and relative strengths might in principle affect the resulting frequency-dependent ellipticity. Hence, it needs to be verified that the polarization fan harmonics are not washed-out by volume averaging.

Here we demonstrate that the polarization-fan property of the high harmonics in our geometry is largely insensitive to variations in the driver's intensities and intensity ratio. Figure S1 presents the numerical results of a bichromatic scheme with $\delta = 0.1$ where the two drivers have equal intensities: $I_{0,1} = I_{0,2} = 7.02 \cdot 10^{12} W/cm^2$ (red curve), $I_{0,1} = I_{0,2} = 6.32 \cdot 10^{13} W/cm^2$ (green) and $I_{0,1} = I_{0,2} = 1.12 \cdot 10^{14} W/cm^2$ (blue). The HGS (panel a) and elipticity-helicity product (b) are shown for the 9th harmonic, and panels (c-d) show similar results for the 15th harmonic. In panels (e-h) the intensity of the first driver is fixed at $I_{0,1} = 1.12 \cdot 10^{14} W/cm^2$, and the second driver's intensity was set at 4 different values, given by intensity ratios $I_{0,2}/I_{0,1} = 0.25, 0.5625, 1, 1.5625$. The HGS (panel a) and elipticity-helicity product (b) for the 9th harmonic are shown on panels (e,f) respectively and similar results for the 15th harmonic are shown in panels (g,h). Clearly, the single-atom ellipticity remains largely unchanged upon changing the absolute or relative intensities of the two drivers. We verified that this is also the case for the spectral phase (both components in the polarization plane). All in all, our calculations indicate that the polarization-fan is not washed-out by volume averaging and maintains the single-atom polarization property through propagation.

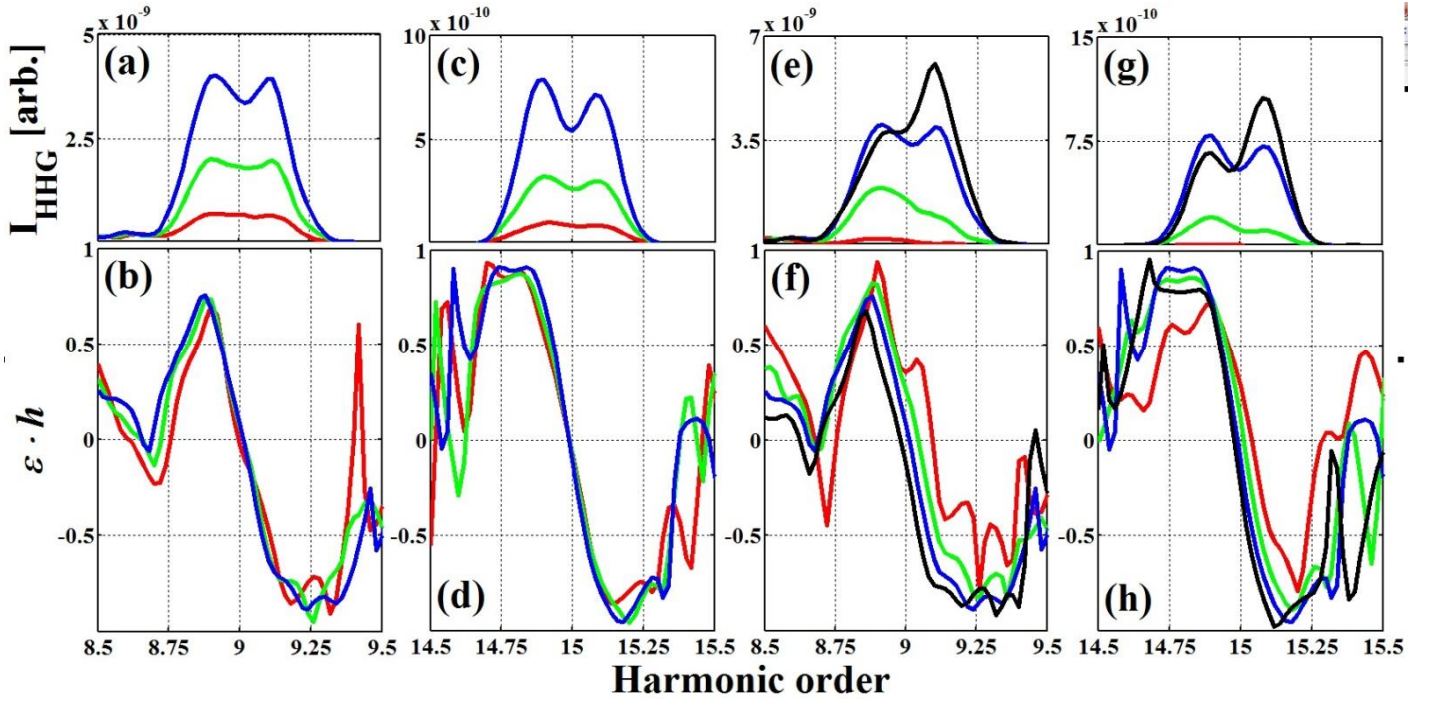

Figure S1: Robustness of polarization-fan harmonics. HGS and ellipticity-helicity product for a $(0.9\omega, 1.1\omega)$ scheme ($\delta = 0.1$) where the two pulses are circularly-polarized and counter-rotating ($\alpha_1=\alpha_2=45^0$). The two pulses have the same intensities, given by $I_{0,1} = I_{0,2} = 7.02 \cdot 10^{12} W/cm^2$ (red curve), $6.32 \cdot 10^{13} W/cm^2$ (green) and $1.12 \cdot 10^{14} W/cm^2$ (blue). (a) HGS of the 9th harmonic. (b) elipticity-helicity product. (c,d) same as (a,b) but for the 15th harmonic. In panels (e-h) the intensity of the first pulse is $I_{0,1} = 1.12 \cdot 10^{14} W/cm^2$, and the second pulse has 4 possible intensities, given by the intensity ratios $I_{0,2}/I_{0,1} = 0.25, 0.5625, 1, 1.5625$ (red, green, blue, black curves, respectively). (e) HGS of the 9th harmonic. (f) elipticity-helicity product. (g,h) same as (e,f) but for the 15th harmonic.